\documentclass[12pt,letter,useAMS,natbib_jrm]{mn2e} 
\usepackage{graphicx,amssymb} 
\usepackage{txfonts} 
\usepackage{natbib_jrm}

\bibpunct[, ]{(}{)}{;}{a}{}{,}

\def \aj {AJ} 
\def \mnras {MNRAS} 
\def \pasp {PASP} 
\def \apj {ApJ} 
 
\def \apjl {ApJL} 
\def \aap {A\&A} 
 
\def \nat {Nature} 
\def \araa {ARAA}

\def\lesssim{\mathrel{\hbox{\rlap{\hbox{\lower4pt\hbox{$\sim$}}}\hbox{$<$}}}} 
\def\gtrsim{\mathrel{\hbox{\rlap{\hbox{\lower4pt\hbox{$\sim$}}}\hbox{$>$}}}}

 \long
\def\symbolfootnote[#1]#2{
\begingroup
\def\thefootnote{\fnsymbol{footnote}}\footnote[#1]{#2}
\endgroup} 
\begin{document}

\title[Progenitor of SN 2011dh]{Did the progenitor of SN~2011dh have a binary companion?
\thanks{Based on observations made with the NASA/ESA Hubble Space Telescope, which is operated by the Association of Universities for Research in Astronomy, Inc., under NASA contract NAS 5-26555. These observations are associated with program GO-13433 and GO-13345.}} 
\author[Maund et al.]{\parbox[t]{\textwidth}{\raggedright J.~R. Maund$^{1,2}$, I. Arcavi $^{3,4}$, M. Ergon$^{5}$, J.~J. Eldridge$^{6}$, C. Georgy$^{7}$, S.~B. Cenko$^{8}$, A. Horesh$^{9}$, R.~G. Izzard $^{10}$\& R. Stancliffe$^{11}$} 
\vspace*{6pt}\\
$^{1}\,$ The Department of Physics and Astronomy, Hicks Building, Hounsfield Road, Sheffield, S3 7RH, U.K.\\
$^{2}\,$ Royal Society Research Fellow\\
$^{3}\,$ Las Cumbres Observatory Global Telescope Network, 6740 Cortona Dr., Suite 102, Goleta, CA 93117, USA\\
$^{4}\,$ Kavli Institute for Theoretical Physics, University of California, Santa Barbara, CA 93106, USA \\
$^{5}\,$ Stockholm University, The Oskar Klein Centre, AlbaNova, SE-106 91 Stockholm, Sweden\\ 
$^{6}\,$ Department of Physics, University of Auckland, Private Bag 92019, Auckland, New Zealand\\
$^{7}\,$ Astrophysics group, EPSAM, Keele University, Lennard-Jones Labs, Keele, ST5 5BG, UK\\
$^{8}\,$ Goddard Space Flight Center, Mail Code 661, Greenbelt, MD 20771, USA\\
$^{9}\,$ Department of Astronomy, California Institute of Technology, MC 249-17, 1200 East California Boulevard, Pasadena, CA 91125, USA\\
$^{10}\, $ Institute of Astronomy, University of Cambridge, Madingley Road, Cambridge, CB3 0HA, UK\\
$^{11}\,$ Argelander-Institut f\"{u}r Astronomie, Auf dem H\"{u}gel 71, D-53121 Bonn, Germany }
\maketitle

\begin{abstract}
We present late-time Hubble Space Telescope (HST) ultraviolet (UV) and optical observations of the site of SN~2011dh in the galaxy M51, $\sim 1164\,$ days post-explosion.  At the SN location, we observe a point source that is visible at all wavelengths, that is significantly fainter than the spectral energy distribution (SED) of the Yellow Supergiant progenitor observed prior to explosion. The previously reported photometry of the progenitor is, therefore, completely unaffected by any sources that may persist at the SN location after explosion.   In comparison with the previously reported late-time photometric evolution of SN~2011dh, we find that the light curve has plateaued at all wavelengths.
The SED of the late-time source is clearly inconsistent with a SED of stellar origin.  Although the SED is bright at UV wavelengths, there is no strong evidence that the late-time luminosity originates solely from a stellar source corresponding to the binary companion, although a partial
contribution to the observed UV flux from a companion star can not be ruled out.
\end{abstract}

\begin{keywords}
supernovae:general -- supernovae:individual (2011dh) 
\end{keywords}

\section{Introduction} 
\label{sec:intro}
Most massive stars with $M_{ZAMS} > 8M_{\odot}$ are expected to end their lives as core-collapse supernovae (SNe) \citep{2009ARA&A..47...63S} . While there has been major success in the direct identification of the red supergiant progenitors of the H-rich Type IIP SNe, attempts to identify the progenitors of the H-deficient Type Ibc SNe have been less conclusive \citep{2013MNRAS.436..774E}.  There are predicted to be two distinct progenitor channels for these SNe: extremely massive stars ($>30-40M_{\odot}$) which lose their hydrogen envelopes through intense stellar winds, and lower mass stars undergoing a binary interaction. \citet{2013MNRAS.436..774E} showed that, given the available pre-explosion detection limits for nearby Type Ibc SNe, the underlying population of progenitors was required to come from a mixture of the two progenitor channels.  Evidence from the sole detection of the progenitor of a Type Ib SN, iPTF~13bvn, suggests that a lower mass binary progenitor may be responsible, rather than a single massive star \citep{2013ApJ...775L...7C,2013A&A...558L...1G,2014A&A...565A.114F,2014AJ....148...68B,2015MNRAS.446.2689E}.  

SNe of Type IIb initially display hydrogen in their early spectra; however, with time these H features grow weaker until the SN transitions to a H-deficient Type Ib SN. Type IIb SNe may, therefore, be considered ``transition'' SNe connecting the H-rich and H-poor SNe \citep{1993ApJ...415L.103F, fili97}. It is thought that the progenitors of Type IIb SNe have been stripped of all but $\sim 0.1M_{\odot}$ of their H envelope \citep{2014ARA&A..52..487S}, highlighting the importance of mass loss, either through winds or binary interaction to the origin of the subtypes in the SN classification scheme \citep{nombin96}.  Unlike their completely H-deficient cousins, however, there has been significant success in the identification of the progenitors stars of Type IIb SNe in pre-explosion observations \citep{alder93j, 2008MNRAS.391L...5C, 2011ApJ...739L..37M,2011ApJ...741L..28V,2014AJ....147...37V}.

The small amount of residual hydrogen on the progenitor means that the degree of mass loss prior to explosion required to reproduce both the progenitor and SN characteristics must be very finely tuned in theoretical models. This makes the progenitors of Type IIb SNe exquisite probes of stellar mass loss through both binary interactions and winds.  As demonstrated by \citet{2009Sci...324..486M}, late-time observations of the sites of SNe, for which the progenitors have been identified in fortuitous pre-explosion images, allows for an enhanced analysis of the progenitor by studying what is left over after the explosion.  This approach permits confirmation of the original identification of the progenitor candidate (through it's disappearance), the search for any remaining stellar components (a binary companion) and enhanced photometric accuracy (through the use of image subtraction techniques).  Here, using late-time Hubble Space Telescope (HST) observations, we apply these techniques to the study of the progenitor of SN~2011dh.

SN 2011dh was discovered on 2011 May 31 in the galaxy M51 \citep{2011arXiv1106.3551A}.  A yellow stellar source was detected at the SN location in pre-explosion HST images, and interpreted as a supergiant by \citet{2011ApJ...739L..37M} and \citet{2011ApJ...741L..28V}.  The SN exhibited rapid temperature evolution, however, during the shock cooling phase \citep{2011arXiv1106.3551A} which indicated the possible explosion of a compact progenitor rather than an extended supergiant.  Late-time observations by  \citet{2013ATel.4850....1V} demonstrated the disappearance of the yellow source in late-time observations, confirming the conclusion of \citeauthor{2011ApJ...739L..37M} that the progenitor of SN~2011dh was a Yellow Supergiant (YSG).  Theoretical models by \citet{2012ApJ...757...31B} and later \citet{2014ApJ...788..193N} were able to reconcile the SN shock cooling behaviour with the extended progenitor observations, concluding that the progenitor had a compact core with a very low-mass extended H-envelope.  In keeping with the study of \citeauthor{2011ApJ...739L..37M}, we adopt a distance to M51 of $7.1 \pm 1.2\,\mathrm{Mpc}$ distance to M51 \citep{2006MNRAS.372.1735T} and a foreground reddening of $E(B-V) = 0.032$ \citep{2011ApJ...737..103S}.

\section{Observations} 
\label{sec:obs} 

\begin{table*}
\caption{\label{tab:obs:11dh} Pre-explosion and late-time HST observations of the site of SN~2011dh in M51} 
\begin{tabular}
{lccccl} \hline\hline Date & Filter & Instrument & Exposure & Pixel Scale & Program\\
(UT) & & & Time(s) & arcsec &\\
\hline \multicolumn{6}{c}{{\bf Pre-explosion}}\\
\hline 21 Jan 2005 & $F435W$ & ACS/WFC & 2720 & 0.025 & 10452$^{1}$\\
21 Jan 2005 & $F555W$ & ACS/WFC & 1360 & 0.025 & 10452\\
21 Jan 2005 & $F658N$ & ACS/WFC & 2720 & 0.025 & 10452\\
21 Jan 2005 & $F814W$ & ACS/WFC & 1360 & 0.025 & 10452\\
13 Nov 2005 & $F336W$ & WFPC2/WF2 & 2600 & 0.1 & 10501$^{2}$\\
\hline \multicolumn{6}{c}{{\bf Late-time}}\\
\hline\\ 
07 Aug 2014 & $F225W$ & WFC3/UVIS & 3772 & 0.02& 13433$^{3}$\\
07 Aug 2014 & $F336W$ & WFC3/UVIS & 1784 & 0.02 &13433\\
10 Aug 2014 & $F435W$ & ACS/WFC & 1072 & 0.025 & 13433\\
10 Aug 2014 & $F555W$ & ACS/WFC & 1232 &0.025 & 13433\\
10 Aug 2014 & $F814W$ & ACS/WFC & 2176 & 0.025 & 13433\\

\hline\hline 
\end{tabular}
\\
$^{1}$ PI: S. Beckwith, $^{2}$ PI: R. Chandar, $^{3}$ PI: J. Maund 
\end{table*}

Details of the pre-explosion and late-time HST observations of the site of SN~2011dh are presented in Table \ref{tab:obs:11dh}. The late-time WFC3/UVIS and ACS/WFC observations were acquired 1164 and 1167 days post-explosion, respectively \citep{2014A&A...562A..17E}.  The pre-explosion observations of the site of SN~2011dh were acquired with the HST Advanced Camera for Surveys (ACS) Wide Field Channel (WFC) and Wide Field Planetary Camera 2 (WFPC2). Details of the pre-explosion HST observations used here (and their reduction and analysis) are presented by \citet{2011ApJ...739L..37M}. The late-time observations of the site of SN~2011dh were conducted in the optical/near-infrared with the ACS/WFC and in the ultraviolet with the Wide Field Camera 3 (WFC3) Ultraviolet and Visible Channel (UVIS), as part of programme GO-13433 (PI Maund) shared with programme GO-13345 (PI Arcavi). The filters utilised for the late-time observations were specifically selected to match the filters of the available pre-explosion observations as closely as possible. The late-time observations for each filter, with both instruments, were acquired using a sequence of four dithered exposures, permitting the rejection of cosmic rays and hot pixels and also providing finer sampling of the point spread function (PSF). The dithered exposures were combined using the {\it astrodrizzle} package, running under {\sc PyRAF}\footnote{STSDAS and PyRAF are products of the Space Telescope Science Institute, which is operated by AURA for NASA}, with final pixel scales of $\mathrm{0.02\, arcsec\,px^{-1}}$ for the WFC3/UVIS observations and $\mathrm{0.025\,arcsec\, px^{-1}}$ for the ACS/WFC observations. Photometry of the late-time observations was conducted using DOLPHOT\footnote{http://americano.dolphinsim.com/dolphot/} \citep{dolphhstphot} with the specific ACS and WFC3 modules.

The position of SN~2011dh on the late-time $F555W$ image was determined with respect to the pre-explosion source, previously identified by \citet{2011ApJ...739L..37M} in the pre-explosion $F555W$ image, to within $0.016\arcsec$ ($N = 23$).  The position of the SN on the late-time WFC3/UVIS $F336W$ image was determined using a geometric transformation calculated between that image and the late-time ACS/WFC $F555W$ image, to within an uncertainty of $0.018\arcsec$ ($N = 20$). The {\sc PyRAF} task {\it crosscor} was used to the refine offset in the registration between the late-time observations that were, nominally, acquired at the same pointing.\\

\section{Results \& Analysis} 
\label{sec:res} 
\begin{figure*}
\includegraphics[width=17.5cm]{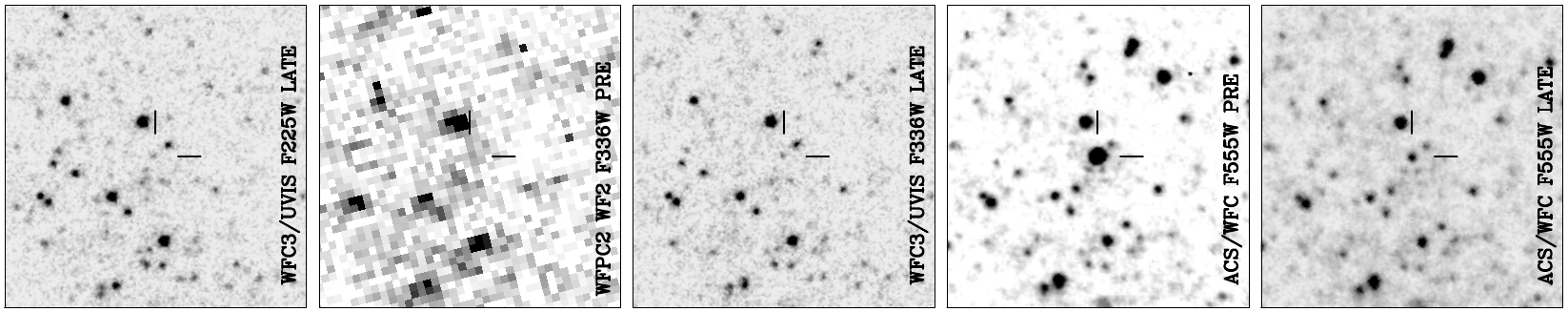} \caption{The site of SN~2011dh observed prior to explosion and at late-times with HST. Each image panel is centred on the SN position (indicated by the cross hairs), with dimensions $4\,\mathrm{arcsec} \times 4\,\mathrm{arcsec}$ and is oriented such that North is up and East is to the left. The panels are, from left to right: Late-time WFC3/UVIS $F225W$ observation; Pre-explosion WFPC2/WF2 $F336W$ observation; Late-time WFC3/UVIS $F336W$ observation; Pre-explosion ACS/WFC $F555W$ observation; and late-time ACS/WFC $F555W$ observation.} 
\label{fig:obs:11dh:panel} 
\end{figure*}

\begin{figure}
\includegraphics[width=8cm]{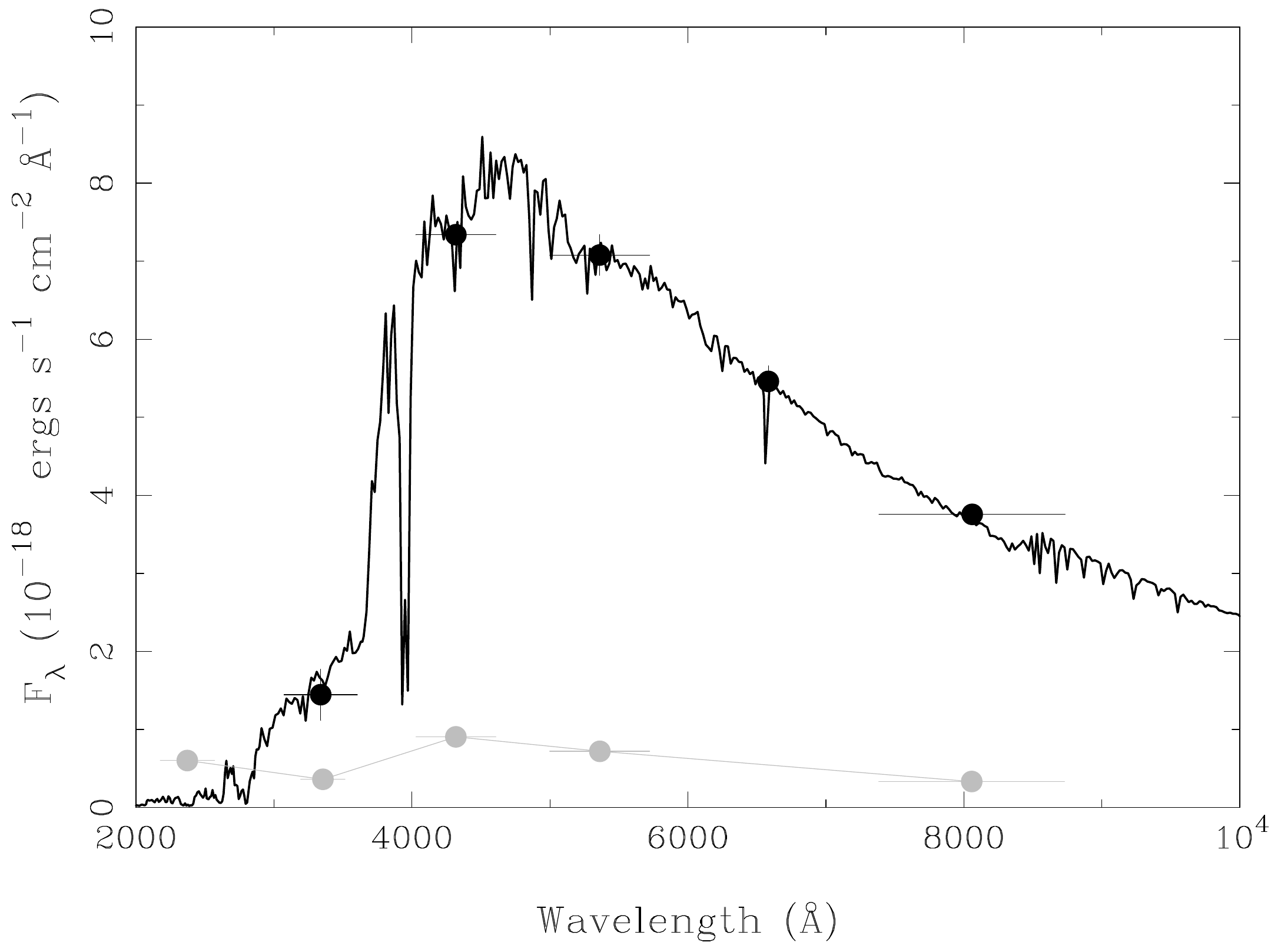} 
\caption{The observed SED of the pre-explosion (heavy points) and late-time (light points) sources at the position of SN~2011dh. Overlaid is a \citet{2004astro.ph..5087C} ATLAS9 synthetic spectrum ($T_{eff}=6000K$) appropriate for the best fit solution for the pre-explosion photometry derived by \citet{2011ApJ...739L..37M}.} 
\label{fig:obs:11dh:sed} 
\end{figure}

A selection of the pre-explosion and late-time observations of the site of SN~2011dh are presented in Figure \ref{fig:obs:11dh:panel}.  In all the late-time observations, from the ultraviolet to the near-infrared, a source is clearly recovered at the SN position. The late-time source is clearly resolved from nearby sources and DOLPHOT PSF-fitting photometry reported the shape of the source (evaluated using the sharpness and $\chi^{2}$ values) to be consistent with a single point source in all bands. Photometry of the late-time source, in comparison with the pre-explosion observations, is presented in Table \ref{tab:res:11dh:phot}. The photometry of the late-time source shows that it is significantly fainter than the pre-explosion in all of the shared photometric bands (see Fig. \ref{fig:obs:11dh:sed}); confirming the previous observations of \citet{2013ApJ...772L..32V} and \citet{2014A&A...562A..17E} that the YSG observed at the SN position prior to explosion is no longer present and, hence, was the progenitor. The relative faintness of the late-time source also implies that any remaining component of the progenitor system (i.e. a binary companion) was not a major contributor to the brightness of the pre-explosion source, in particular in the $F336W$ band (where the late-time flux is lower than the observed pre-explosion flux by a factor of $\sim 4$). This implies that the analysis of the pre-explosion observation of the YSG progenitor, presented by \citet{2011ApJ...739L..37M}, is still valid: $\log (L/L_{\odot}) = 4.92 \pm 0.20$, $T_{eff} = 6000 \pm 280K$ and subject to $E(B-V) = 0 - 0.032$ mags.  In all the late-time observations we see no evidence for spatially resolved light echoes in the vicinity of SN~2011dh, which is consistent with the low degree of reddening reported to SN~2011dh itself. 

\begin{figure}
\includegraphics[width = 8cm]{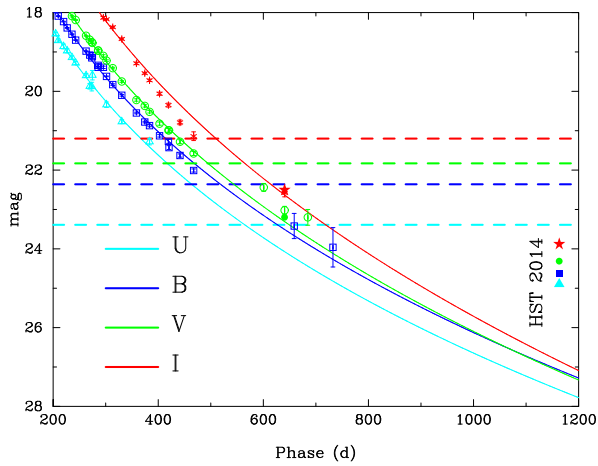} \caption{Late-time photometry of SN~2011dh from \citet{2014arXiv1408.0731E} and HST (from 2013 and 2014; solid symbols).  Solid lines indicate the power law fits to ground based and HST observations prior to our 2014 observations.  Horizontal dashed lines indicate the brightness of the progenitor in the pre-explosion observations.} 
\label{fig:res:11dh:lc}
\end{figure}

Utilising our own SED fitting code, we attempted to fit the SED of the late-time source with ATLAS 9 model SEDs \citep{2004astro.ph..5087C} for single supergiant stars. Using the reddening reported for SN~2011dh as a prior, the most likely solution yielded $\chi^{2}=92.4$, implying that the SED of the late-time source is inconsistent with a stellar source.  Given the late-time photometric evolution of SN~2011dh reported up to $\sim$350d by \citet{2013MNRAS.436.3614S} and $\sim$700d by \citet{2014arXiv1408.0731E}, the shape of the SED suggests a significant contribution from the still bright SN. Using the late-time photometric observations reported by \citet{2014arXiv1408.0731E}, with HST photometry from 2013 \citep{2013ApJ...772L..32V}, we find that the behaviour of SN~2011dh at epochs prior to our HST observations were inconsistent with a linear decay and are, rather, better approximated by a power law (of the form $m = \alpha t^{\beta}$; see Fig. \ref{fig:res:11dh:lc}).   The most recent late-time HST photometric measurements reported here are, however, even inconsistent with these earlier light curve trends and Figure \ref{fig:res:11dh:lc} shows that the light curve has flattened across all wavelengths.

\begin{figure}
\includegraphics[width=8cm]{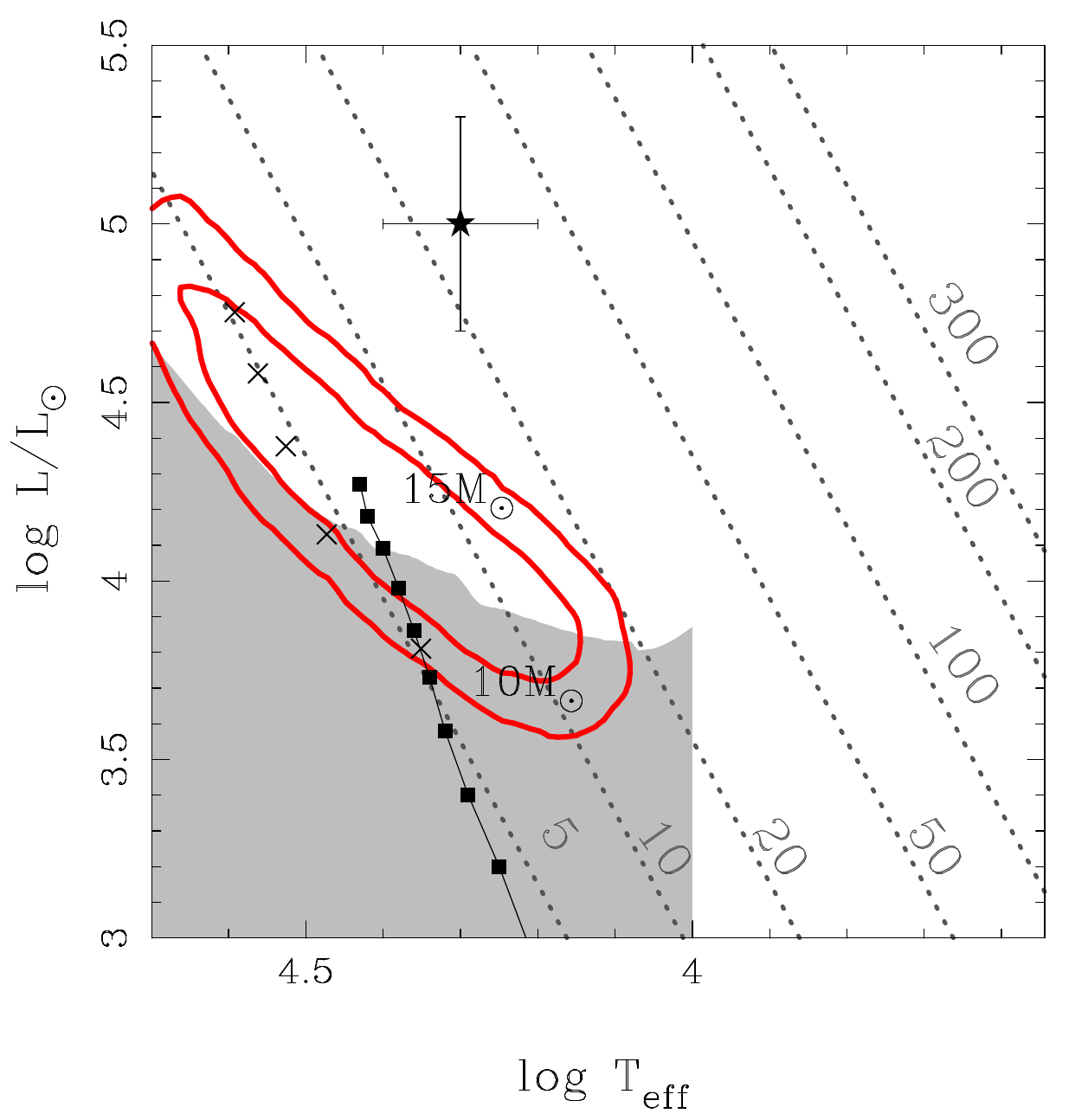} \caption{Hertzsprung-Russell diagram showing the permitted locations for a possible binary companion, given the late-time photometry, under the assumptions: 1) all UV flux arises from the companion (red contours) and 2) only the $F225W$ flux arises from the companion (grey shaded region). The upper limit of the grey shaded region corresponds to position on the HRD where the probability of detection of such a star is 50\%. The starred point corresponds to the position of the companion star identified for the progenitor of Type IIb SN~1993J \citep{maund93j}.   The positions of possible binary companions predicted by \citet{2013ApJ...762...74B} are indicated by the crosses ($\times$) and the main sequence for stars with $M_{ZAMS} \leq 15M_{\odot}$, for integer masses from the STARS stellar evolution code \citep{eld04}, are shown by the filled squares ($\blacksquare$).  The dotted lines indicate lines of constant radius (as labelled in solar radii).} 
\label{fig:res:11dh:hrd} 
\end{figure}

In light of the still significant late-time contribution of the SN at optical wavelengths, it is difficult to place specific constraints on the presence of a binary companion that also contributes to the late-time SED. Under the assumption that the SN component is weakest in the UV, while a companion star's SED might peak in UV, we may attempt to place constraints on the allowed parameters for the companion given these observations. We consider two scenarios for the origin of the late-time UV flux: 1) all of the observed $F225W$ and $F336W$ flux arises from a companion star; and 2) all of the $F225W$ flux and a portion of the $F336W$ flux arises from a companion. Given these assumptions, the corresponding permitted regions for the companion star on the Hertzsprung-Russell diagram (HRD) are presented on Fig. \ref{fig:res:11dh:hrd}. For the second scenario, we utilised the ACS/WFC $F435W$ photometry as a further constraint on the temperature, as stars cooler than $T\sim 10\,000K$ would be brighter than the observed SED at that wavelength.   Under the first assumption, due to correlations between luminosity and temperature, all of the \citet{2013ApJ...762...74B} models for the companions, covering a range of mass accretion efficiencies, fit the observed UV photometry.  Under the second assumption, only the binary companion models with the smallest mass accretion efficiencies ($\beta \leq 0.25$) or main sequence companions with $M \leq 13M_{\odot}$ are allowed.

\begin{table}
\caption{\label{tab:res:11dh:phot} Photometry of the sources at the position of SN~2011dh in pre-explosion and late-time HST observations} 
\begin{tabular}
{ccc} \hline\hline Filter & Pre-explosion & Late-time \\
& (mags) & (mags) \\
\hline F225W & $\cdots$ & 24.61(0.11) \\
F336W & 23.39(0.25)& 24.89(0.11) \\
F435W & 22.36(0.02)& 24.63(0.04) \\
F555W & 21.83(0.04)& 24.31(0.03) \\
F658N & 21.28(0.04)& $\cdots$ \\
F814W & 21.20(0.03)& 23.83(0.02) \\
\hline\hline 
\end{tabular}
\end{table}
\section{Discussion and Conclusions} 
\label{sec:disc}
\begin{figure*}
\includegraphics[width = 11cm, angle = 270]{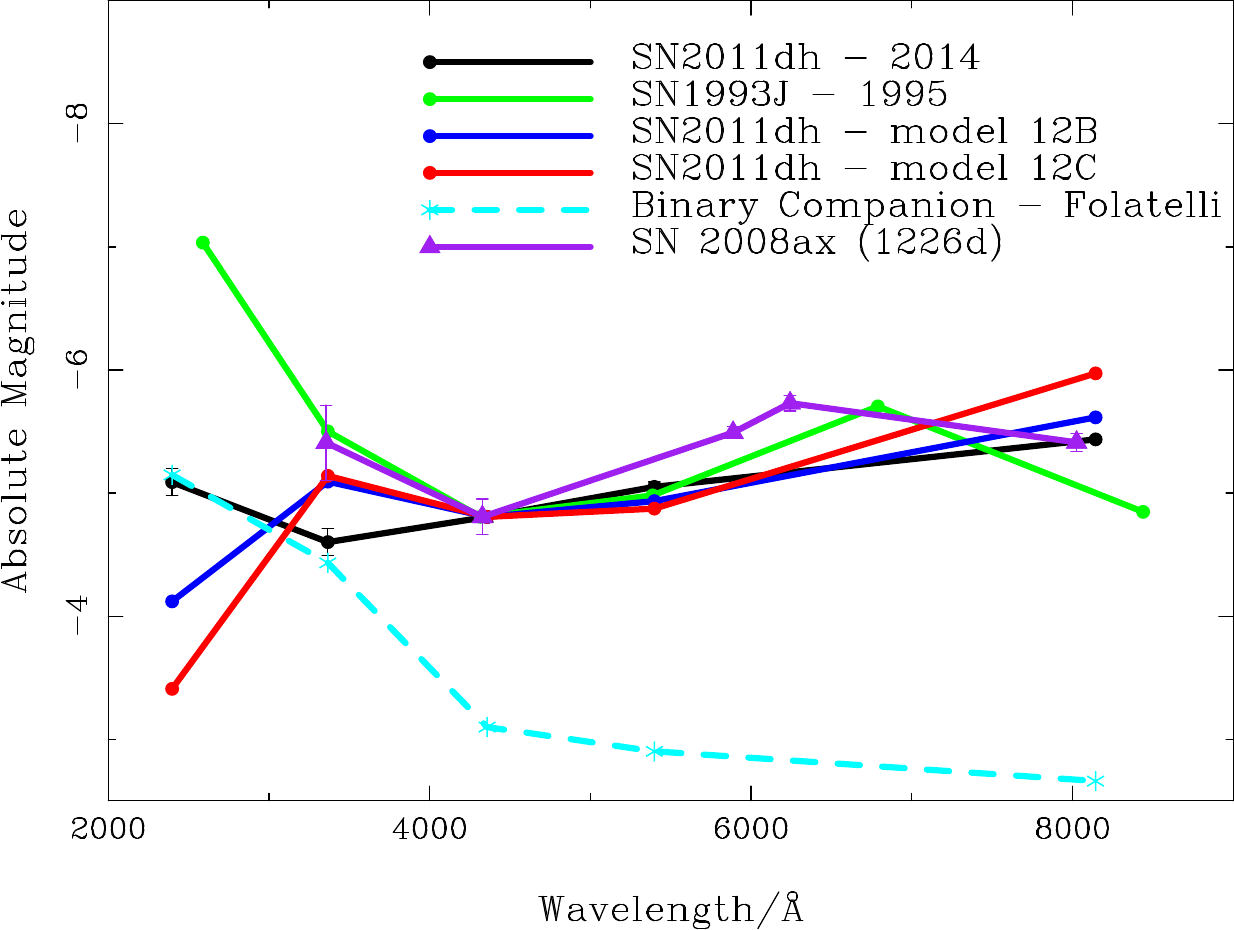}
\caption{Theoretical and observed late-time SEDs, in units of absolute Vega magnitudes, for Type IIb SNe compared to the observed late-time SED of SN 2011dh.  The SEDs are normalised to the brightness of the late-time source recovered at the position of SN~2011dh, as observed in the ACS/WFC $F435W$ band.  Also overlaid are the SED of the binary companion proposed by \citet{2014ApJ...793L..22F} and the observed SED of SN 2008ax at 1226d (corrected for $E(B-V) = 0.4$; Maund et al., 2015).}
\label{fig:disc:11dh:sed}
\end{figure*}

%
%
Previously, using the subset of late-time $F225W$ and $F336W$ observations presented here, \citet{2014ApJ...793L..22F} claimed the detection of a hot, compact companion as predicted by \citet{2013ApJ...762...74B}.  In the course of their analysis, however, \citeauthor{2014ApJ...793L..22F} had to make the key assumption that the UV flux {\it only} originated from the binary companion, and then were only able to constrain the properties using two photometric data points.  A fundamental test of this hypothesis would be the decreasing flux at redder wavelengths, which is clearly at odds with the observed non-stellar optical SED (see Fig. \ref{fig:disc:11dh:sed}).  The late-time brightness of SN~2011dh at optical wavelengths is unexpected given the late-time evolution reported by \citet{2013MNRAS.436.3614S} and \citet{2014arXiv1408.0731E}, but also raises serious concerns about the degree of the SN contribution at UV wavelengths, specifically where \citeauthor{2014ApJ...793L..22F} have claimed to observe the binary companion.

\citet{2014arXiv1408.0731E} discuss the late time evolution of SN 2011dh and find that the radioactive energy deposition is likely insufficient to reproduce the late time (600-750 days) optical luminosity. "Freeze-out" in the helium envelope  could also be a possible additional energy source. \citet{2014arXiv1408.0731E} also report strong flattening of the observed mid-infrared (MIR) lightcurve after 750 days.  If we use the optical decline rate before 750 days and the MIR decline rate thereafter, we find a drop of 1.0 mag between 640 and 1166 days, which is in good agreement with the observed drop of 1.1-1.2 mag in the late-time (2014) F555W and F814W photometry and suggests that the flattening of the light curve is relatively wavelength independent (see Fig. \ref{fig:res:11dh:lc}).

In Fig. \ref{fig:disc:11dh:sed} we show the SEDs for the steady-state non-Local Thermodynamic Equilibrium models 12C and 12B of \citet{2015A&A...573A..12J} at 500 days, who provide examples of how the SED of the SN would look in the radioactively powered scenario. These models only differ in the treatment of positron trapping (local or non-local), and are able to reproduce the lightcurve \citep{2014arXiv1408.0731E} and spectra \citep{2015A&A...573A..12J} of SN 2011dh over the period of 100-500 days. We also show the observed SED of SN 1993J at 674 days \citep{2002PASP..114.1322V}, which provides an example of how the SED would look like if there was a contribution from the strong interaction of the ejecta with the circumstellar medium (CSM). As seen in Fig. \ref{fig:disc:11dh:sed}, the shape of the observed SED in the optical is consistent with what could be expected for the SN in the radioactively powered scenario, but does not explain the UV portion of the SED (and could, possibly, require a binary companion).  This requires, however, that there is no contribution due to the any late interaction with the CSM and that the F225W-F336W colour for the freeze-out powered scenario is not considerably bluer than for the radioactively powered scenario. As discussed in \citet{2014arXiv1408.0731E}, there are no clear signs of significant CSM interaction in the spectrum of SN~2011dh acquired at 678 days presented by \citet{2013MNRAS.436.3614S}, which disfavours the CSM powered scenario. While an investigation of how the SED would look in the freeze-out powered scenario is outside the scope of this paper, at this point we can not rule out that the observed flux in the F225W filter, and any of the other filters, originates solely from the SN. If the optical flux is dominated by the SN at 1164 days, the radioactively powered scenario can be ruled out, as the optical luminosity alone would then be 1.9 mag higher than the bolometric luminosity in models 12C and 12B.

To consider the role of CSM interaction in the behaviour of the late-time lightcurve of SN~2011dh, we obtained target-of-opportunity X-ray observations at the position of SN 2011dh with the X-ray telescope \citep[XRT][]{2004ApJ...611.1005G} on-board the \textit{Swift} Gamma-Ray Burst Explorer \citep{2004ApJ...611.1005G} on 2014 Sep 17 (4.9\,ks) and 2014 Sep 24 (4.9\,ks).   While generally following standard algorithm
for automated analysis of XRT data (e.g., \citealt{2009MNRAS.397.1177E}), we manually 
selected background regions due to the complex nature of diffuse
and/or unresolved emission from M51\footnote{Note that we verified that
this diffuse component is unrelated to SN 2011dh by examining 
\textit{Swift}-XRT observations of the field from 2005.}  We find 
no evidence for significant point source emission (i.e., above and beyond 
the underlying background) at the position of SN 2011dh, down to a $3\sigma$ limit of $2\times10^{-13}$\,erg\,s$^{-1}$\,cm$^{-2}$, which corresponds to a luminosity of $\sim1.22\times10^{-39}$\,erg\,s$^{-1}$.

We also obtained late time radio observations of SN\,2011dh with the Jansky Very Large Array (VLA). The observations were performed on October 18, 2014 UT at central frequencies of $6.1$\,GHz (C-band) and $22$\,GHz (K-band) using 3C286 as a flux calibrator and J1349+5341 as phase calibrator. We detect weak radio emission from the SN at flux levels of $459\pm 93$\,$\mu$Jy and $109\pm 17$\,$\mu$Jy at C-band and K-band, respectively. Past radio observations, from day $\sim 4$ until day $\sim 90$ after explosion, have already shown radio emission thus indicating interaction of the SN ejecta with CSM
\citep{2012ApJ...750L..40K,2012ApJ...752...78S,2013MNRAS.436.1258H}.  Our late time radio measurements are well within the expectations based on the earlier measurements, assuming that the CSM has an extended wind-like structure. The lack of excess in radio emission thus suggests that there is no additional interaction with an enhanced CSM. 

While the X-ray and radio observations rule out strong ongoing CSM interaction at the time of the late-time HST observations, portions of the UV and optical SED could be enhanced by emission lines generated in some low level interaction.  Lines such as {\sc C ii} $\lambda\lambda 2324,2325$ and [Ne {\sc iii}] $\lambda\lambda 3868, 3869$ \citep{1999AJ....117..725F} could contribute flux to the observed F225W and F336W photometry.  Ultimately a UV late-time spectrum of SN~2011dh, which is unfortunately unfeasible with HST, would be required to establish if such emission lines are present, without a commensurate signal for CSM interaction in the X-ray and radio regimes.  While earlier CSM interaction is ruled out by the spectra presented by \citet{2013MNRAS.436.3614S}, without similar spectra at epochs comparable to our late-time HST observations it is difficult to assess whether any interaction might have started in the intervening period between those observations and our 2014 observations when the light curve appears to have flattened (which could mark a period of earlier enhanced mass loss in the history of the progenitor).     

Alternatively, the late-time flux could arise from early SN flux scattered into the line of sight by dust, forming a light echo \citep{2011ApJ...732....2R}.  We note, however, that in all the late-time observations SN~2011dh appears point-like, which would require the scattering dust to have a very specific arrangement, lying behind the SN and exactly along the line of sight, which would be unlikely to occur by chance.  In addition, given the low reddening estimated towards SN~2011dh there is no evidence for large amounts of dust in the SN locality.  We therefore discount the possibility of a light echo being the origin of the observed late-time brightness of SN~2011dh.

Based on the late-time HST observations presented here, there is, as yet, no conclusive evidence for the recovery of a binary companion associated with SN~2011dh.  The presence of significant late-time flux at optical wavelengths for SN 2011dh suggests that the origin of the UV flux is highly uncertain and not necessarily attributable to a binary companion.  HST observations of SN 2008ax at a similar epoch in its evolution (Maund et al., in prep.), exhibited a similar rise at UV wavelengths (see Fig. \ref{fig:disc:11dh:sed}), however there is no comparable observation with shorter wavelength filter to confirm that this rise in the SED continues in the UV as observed for SN~2011dh.  In the case of SN~2008ax, even later HST observations confirmed the disappearance of the progenitor star, with no stellar residual at the SN location, corresponding to a binary companion, at UV or optical wavelengths.  Alternatively, given the late-time brightness of SN~2011dh, a binary companion could be still be "hidden" by the SN, in which case its luminosity and, hence, mass would be lower than the optimistic interpretation of the UV observations presented by  \citet{2014ApJ...793L..22F} and its role in the evolution of the progenitor might be diminished.  Constraining the progenitor to a lower mass has important implications for interacting binary evolution models. A lower mass would indicate that the mass transfer responsible for removing the hydrogen
envelope was less efficient than for the case of SN 1993J, with most of the mass lost from the binary system rather than ending up in the companion star. Later, deeper observations once SN~2011dh has faded even more may permit the detection of a lower mass binary companion if it is present. 

A fundamental question raised by these observations, however, is whether a binary companion is a prerequisite for the final appearance of the progenitor of SN~2011dh.   There is significant evidence for the presence of a hot binary companion to the progenitor of SN~1993J \citep{alder93j,maund93j,2014ApJ...790...17F}, the class prototype of Type IIb SNe.  \citet{2012A&A...538L...8G}, however, found that it was possible to produce YSG progenitors through single star evolution through the inclusion of rotation and by enhancing the mass loss for a $M_{ZAMS} = 15M_{\odot}$ star by a factor or 10.  This enhanced mass loss rate is, however, consistent with the observed levels of mass loss for some Red Supergiants \citep{2005A&A...438..273V} and given the measured luminosity of the progenitor of SN~2011dh the expected mass loss rate of $\sim 10^{-5}M_{\odot}\,\mathrm{yr^{-1}}$ \citep{2012A&A...542A..29G} is consistent with the mass loss rate inferred from radio observations of the SN itself \citep{2012ApJ...752...78S}.  In isolation, for a rotating $M_{ZAMS}=18M_{\odot}$ star, \citet{2013A&A...558A.131G} found the progenitor evolved to become a YSG, instead of exploding as a Red Supergiant.  This may suggest that new mass loss rate prescriptions and rotation may make YSGs the natural endpoint for some stars of mass $\sim 13M_{\odot}$, such as the progenitor of SN~2011dh.  It should be noted that the mass-loss rates of RSGs are poorly known; even for stars of similar luminosities mass-loss rate determinations from the literature \citep{2005A&A...438..273V,2011A&A...526A.156M} vary over several orders of magnitudes. The reason for the scatter is unclear, which could be due to unseen companions, eruptive episodes of mass loss or a yet to be identified physical process, but could lead to stars of similar initial mass evolving very differently towards the ends of their lives.

While mass transfer in binary systems can be invoked to explain the observed properties of H-poor YSG progenitors, there is a significant increase in the parameter space in which acceptable models might arise.  From the perspective of stellar evolution models, in both the single and binary scenarios, predicting the final temperature of a YSG progenitor requires fine tuning the adopted mass loss rate.  The case of the progenitor of SN~2011dh may serve to highlight the importance of understanding gaps in our knowledge of single star evolution and of how the light curves of SNe at late-times are powered.

\section*{Acknowledgements} The authors are very grateful to Anders Jerkstrand for providing us with access to his late-time model spectra of SN~2011dh.  The research of JRM is supported by a Royal Society University Research Fellowship. C.G. acknowledges support from the European Research Council under the European Unions Seventh Framework Program (FP/2007-2013)/ERC Grant Agreement No. 306901.

\bibliographystyle{apj}

\end{document}